\documentstyle[preprint,aps,12pt]{revtex}

\newcommand{\la}{\langle}
\newcommand{\ra}{\rangle}

\newcommand{\beq}{\begin{eqnarray}}
\newcommand{\eeq}{\end{eqnarray}}

\renewcommand{\Im}{{\rm Im}}
\renewcommand{\ln}{{\rm ln}}

\renewcommand{\Re}{{\rm Re}}

\newcommand{\btem}{\bibitem}

\begin{document}

\preprint{UTHEP-299, April 1995}

\title{ The Pion-Nucleon Coupling Constant\\
  in  QCD Sum Rules}

\author{H. Shiomi$^1$
 and T. Hatsuda$^{1,2}$}

\address{$^1$ Institute of Physics, University
 of Tsukuba, Tsukuba, Ibaraki 305, Japan}

\address{$^2$ Institute for Nuclear Theory,
 NK-12, Univ. of Washington, Seattle, WA 98195, USA}

\date{\today}

\maketitle

\begin{abstract}
 The pion-nucleon coupling constant $g_{\pi N}$  
is studied on the basis
 of  the QCD sum rules. 
 Both the Borel sum rules and the finite energy sum rules 
 for $g_{\pi N}$ are used to examine the
 effects of higher dimensional operators (up to dim. 7) 
 and  $\alpha_s$ corrections in the operator product expansion.
  Agreement with the experimental number is reached 
only
 when
 $S_{\pi}/S_N$ is greater than one,
 where $S_{\pi}$ ($S_N$) is the continuum threshold for
 the $g_{\pi N}$ (nucleon) sum rule.
 
\end{abstract}

\newpage
\setcounter{equation}{0}
\section{Introduction}
 The pion-nucleon coupling
 constant $g_{\pi N}$ is one of the most fundamental quantities
 in hadron physics. $g_{\pi N} = 13.4 \pm 0.1$ is obtained empirically  
 using the $N-N$ scattering data,  $\pi-N$ scattering data and the 
deuteron
properties \cite{EW}.  From the theoretical point of view,
 it is a great challenge to reproduce this value from the 
 first principle namely
 the quantum chromodynamics (QCD).  So far,
 there have been two attempts;
 one is based on the lattice QCD simulations (see ref.\cite{liu})
 and the other is 
 based on the
 QCD sum rules (QSR) \cite{Shifman,yazaki}.  The latter approach 
for
 $g_{\pi N}$, however,  has not been explored in great detail 
beyond the
 leading order of the operator product expansion (OPE) 
\cite{RRY,yazaki}
  within the authors' knowledge.\footnote{There is a QSR study
 of the axial charge $g_A$, which can be related to $g_{\pi N}$
 assuming the Goldberger-Treiman (GT) relation \cite{henley}.
  In the present paper, however, we will study $g_{\pi N}$ 
directly.}
 The purpose of this paper is to reexamine the problem 
  using the currently available information on the higher 
dimensional
 operators in OPE and the $\alpha_s$ corrections to the
 Wilson coefficients.
 
To study $g_{\pi N}$ in QSR, 
  two methods have been proposed so far: 
 (i) method based on the two point function
 $\la 0 \mid {\rm T} \eta(x) \bar{\eta}(0) \mid \pi \ra$,
 and (ii) method based on the three point function
 $\la 0 \mid {\rm T} \eta(x) \bar{q} i \gamma_5 q(y) \bar{\eta}(0)
  \mid 0 \ra $, where $\eta(x)$ is the nucleon interpolating field.
  We will take the first approach (i) throughout this paper, since
  OPE in (ii) is problematic when the four momentum of the 
 pion becomes soft.
   The lowest non-trivial order of OPE in (i)
 is known to
 give  $g_{\pi N}^{lowest}={ M_N / f_\pi}$ when the
continuum threshold is neglected \cite{yazaki,RRY}.
 Th origin of the 25\% disagreement of $g_{\pi N}^{lowest}$ from 
the Goldberger-Treiman (GT) relation
  $g_{\pi N} ={g_A M_N / f_{\pi}}$ may originate
 either from the higher dimensional operators, $\alpha_s$ 
corrections
 and  the continuum threshold.
 We will examine $g_{\pi N}$ with all these ingredients.

 In section II, we will  examine  QSR for
 the nucleon mass by adopting   OPE up to dimension 7 operators 
with
   $O(\alpha_s)$ corrections, which is 
  essential for the discussions in
  later sections.
 In section III, QSR for $g_{\pi N}$ is studied in close analogy
with that for the nucleon mass.
In section IV,  Borel  analyses for $g_{\pi N}$ 
 are made, and the effect of the $\alpha_s$ corrections, higher 
dimensional
operators and the continuum thresholds are studied.
 Section V is devoted to summary and concluding remarks.


\section{ sum rules for the nucleon }
 Let's consider the two point function,
\beq
\label{correlation}
{\Pi_{\alpha \beta }(q)}  & = & i\int d^4x e^{iq \cdot x} 
\la 0 | T\eta_\alpha (x) \bar{\eta}_\beta (0) |0 \ra  \nonumber \\
   & = & 
 \Pi_1(q) \hat{q}_{\alpha \beta} + \Pi_2(q){\bf 1}_{\alpha \beta},
\eeq
with $\hat{q}\equiv q\cdot\gamma$.
 For interpolating nucleon current, we use the Ioffe current 
\cite{Ioffe}
 \beq
\label{current}
\eta(x)=\epsilon_{abc}(u^a(x)C \gamma_\mu u^b(x))
\gamma_5 \gamma^\mu d^c(x),  
\eeq
where $a$, $b$ and $c$ are color indices and $C$ is 
the charge conjugation  operator.

 The operator product 
expansion (OPE) at  $q^2 \rightarrow - \infty$ has a general form
\beq
\label{OPE}
\int d^4x e^{iq\cdot x} T\eta(x) {\bar \eta}(0) =
  \sum_n C_n(q, \mu, \alpha_s(\mu^2)) O_n (\mu),
\eeq 
where $C_n$ denote the Wilson coefficients and $O_n$ are the local
 gauge invariant operators. They depend on the renormalization 
scale
 $\mu$ which separates the short distance dynamics  in
 $C_n$  and the 
 long distance dynamics  in $O_n$.   
 If one takes the vacuum expectation value of (\ref{OPE}), it can be used 
for 
 the sum rules of the nucleon, while if one takes the vacuum to 
pion
 matrix  element, it can be used for the sum rules of $g_{\pi N}$.
  
The Lorentz structure  of $O_n$ depends on the 
 states one chooses to sandwich (\ref{OPE}).
 For the nucleon sum rules, only the scalar operators contribute.
 OPE up to dimension 7 with  $\alpha_s$
 corrections in the chiral limit can be extracted from  
 refs. \cite{YHHK} and \cite{ovchinnikov};
\beq 
\label{nucleon.ope}
& & \Re \Pi(Q^2)=\left(\Pi^a_1+\Pi^b_1+\Pi^c_1 \right) \hat{q}
+\left(\Pi^d_2+\Pi^e_2 +\Pi^f_2 \right) ,\\
& & \Pi_1^a(Q^2)={-1 \over 64 \pi^4} Q^4 \ln{Q^2 \over \mu^2} 
\left (1+ {71 \over 12}{\alpha_s \over \pi}-{1 \over 2}{\alpha_s 
\over \pi}
\ln{Q^2 \over \mu^2} \right) ,\\
& & \Pi_1^b(Q^2)={-1 \over 32 \pi^2} \la {\alpha_s \over \pi}G^2 \ra
 \ln{Q^2 \over \mu^2} , \\
& & \Pi_1^c(Q^2)= \frac23{\la\bar{u}u\ra^2 \over  Q^2} \left 
\{1-{\alpha_s \over \pi}
\left(\frac13 \ln{Q^2 \over \mu^2}+\frac56 \right) \right \} ,\\
\label{nucleon.ope1}
& & \Pi_2^d(Q^2)={-\la \bar{d}d \ra \over 4\pi^2} Q^2\ln{Q^2 \over 
\mu^2}
  \left(1+\frac32 {\alpha_s \over \pi} \right) ,\\  
\label{nucleon.ope2}
& & \Pi^e_2(Q^2)=0 ,\\
\label{nucleon.ope3}
& &\Pi_2^f(Q^2)={1 \over 18 Q^2} \la {\alpha_s \over \pi} G^2 \ra
\la\bar{d}d\ra ,
\eeq
where $Q^2 \equiv -q^2 \rightarrow \infty$ and 
  $\la \cdot \ra$ denotes the vacuum expectation
 value. The argument of $\alpha_s$ is $\mu^2$ which is not written
  explicitly. The  diagrammatic illustration of $\Pi^a_1 \sim
 \Pi^f_2$ 
  is shown in Fig. 1.

 Several  remarks are in order here.\\
  (a) We take the chiral limit ($m_q=0$) throughout this paper.
 Small $u,d$ quark mass does not change the essential conclusion
 of this paper. \\
 (b) The fact that the Wilson coefficient of the
 dimension 5 operator $g \bar{q} \sigma \cdot G q$ vanishes
  in $\Pi_2$  is a 
 unique property of the Ioffe current \cite{Ioffe}. 
 Since this operator is already $O(g)$,  we
 do not consider the $O(\alpha_s)$ correction to it.\\
 (c) A small discrepancy between the formula in 
 \cite{ovchinnikov} and that in 
 \cite{Jamin} for  $\Pi_2^d$
  has been  recently resolved  (see \cite{AP}) and the 
 final result boils down to
 the form given in the above. \\
 (d) We always assume  the  vacuum saturation
  when evaluating  matrix elements of 
 higher dimensional operators.

The  correlation function 
 (\ref{correlation}) satisfies the standard dispersion relation
\beq
\label{nucl.dispersion}
\Re \Pi_{1,2}(Q^2)&=&{1 \over \pi} \int {  \Im\Pi_{1,2} (s) 
 \over s+Q^2} ds 
+ {\rm subtraction}.
\eeq
 In QSR, $\Re \Pi(Q^2)$ in the left hand side of 
(\ref{nucl.dispersion})
 is calculated by OPE at large $Q^2$ as give in 
(\ref{nucleon.ope}),
 while $\Im \Pi(s)$
in
 the right hand side is  
  parametrized by the nucleon pole and the phenomenological 
 continuum. The pole part reads
\beq
\label{pole}
\hat{q} \Im\Pi_1^{pole}(s) + \Im\Pi_2^{pole}(s) =  \pi \lambda_N^2 
 (\hat{q}+ M_N) \delta(s-M^2_N),
\eeq
 where $\lambda_N$ is defined as $\la 0 | \eta | N \ra = 
 \lambda_N u(p)$ with $u(p)$ being the nucleon Dirac spinor.
  We assume that, when  $s > S_N$,  the hadronic continuum reduces 
to the same form 
  with that obtained by an analytic continuation
 of OPE;
 \beq
\label{cont}
 \Im \Pi_1^{cont}(s)&=& \pi \theta(s-S_N) \left[{s^2 \over 64\pi^4}
 \left \{ 1+{\alpha_s \over \pi}\left (
 \frac{71}{12}-\log({s\over \mu^2}) \right) \right \} +{1 \over 
32\pi^2}
 \la {\alpha_s \over \pi}G^2 \ra \right]  , \\
\label{cont2}
 \Im \Pi_2^{cont}(s)  & =- \pi & \theta(s-S_N){\la \bar{d} d\ra \over 
4\pi^2}
       s \left(1+\frac32 {\alpha_s \over \pi} \right) .
\eeq       
  
 Substituting  (\ref{nucleon.ope}) and (\ref{pole},\ref{cont},\ref{cont2}) 
 into the dispersion relation (\ref{nucl.dispersion}),
  one can generate sum rules for the resonance parameters.
 We will write here the Borel transformed version of 
 the sum rules in which  
 the higher dimensional operators and the effect of the continuum
 can be relatively suppressed. (See Appendix A for useful formula.)
\beq
\label{nucleonQSR1}
4\pi^4 \lambda_N^2 e^{-M_N^2/ M^2}
&=&{M^6 \over 8}E_2(x) \left(1+(\frac{53}{12}+\gamma_E) 
{\alpha_s(M^2) \over
\pi} \right)\nonumber  \\ & &+{b M^2 \over 32}E_0(x)+{a_u^2 \over 
6}
\left(1-(\frac56 -\frac13 \gamma_E ){\alpha_s(M^2) \over  \pi} 
\right) ,  \\
\label{nucleonQSR2}
4\pi^4 \lambda_N^2 M_N e^{-M_N^2/ M^2}
&=&{a_d \over4}M^4 E_1(x) \left(1+\frac32 {\alpha_s(M^2) \over \pi} 
\right) -{a_d b \over 72}.
\eeq
Here $M^2$ is the Borel mass, 
 $E_n=1-(1+x+\cdots+{x^n \over n!})e^{-x}$ with $x={S_N/M^2}$, 
 $\gamma_E$ is the 
Euler constant (0.5772$\cdots$) and 
\beq
 a_q \equiv -4 \pi^2\la\bar{q}q\ra, \ \ \ \ \ \ 
b \equiv 4\pi^2\la{\alpha \over \pi}G^2\ra.
\eeq
 Note that we chose $\mu^2 $ to be $M^2$ which is 
 a typical scale of the system after 
  the Borel transform.  We call eq. (\ref{nucleonQSR1})
 (eq. (\ref{nucleonQSR2}))  as ``even" (``odd") sum rule
 since it contains only even (odd) dimensional operators. 
 As for  the running coupling constant, we use a simplest one-loop
 form,
\beq
\alpha_s(M^2)={4\pi \over 9 \log(M^2/\Lambda^2)},
\eeq
with $\Lambda^2=(0.174GeV)^2$ which is obtained to reproduce
 $\alpha_s(1) \simeq 0.4$ \cite{data}.

  Eq.(\ref{nucleonQSR1}) and Eq. (\ref{nucleonQSR2}) will be 
 used  when we analyse the sum rules
 for $g_{\pi N}$.  One can derive formula for $M_N$ as a function 
of
 $M$ in three
ways; 
 (i) the ratio of  (\ref{nucleonQSR1}) and (\ref{nucleonQSR2}),
(ii) the ratio of  (\ref{nucleonQSR1})
 and its logarithmic derivative with respect to $M^2$ and 
 (iii) the ratio of (\ref{nucleonQSR2})
 and its logarithmic derivative with respect to $M^2$.
 We have made an extensive Borel stability 
 analyses for these three cases.
 We found that the higher dimensional operator and the $\alpha_s$ 
corrections
 improve the Borel stability as well as the prediction for  $M_N$.  However,
there is a wide range of $S_N$ which can  reproduce the
 experimental nucleon mass in the Borel analyses.  
 In  later sections, we will use the finite energy sum rules 
(FESR) to fix $S_N$. 

In Fig.1, shown are the Borel curves for three 
cases (i)-(iii)
 with $S_N = 1.601 GeV^2$  obtained by the FESR with 
 $\la \bar{q}q \ra = -(0.2402GeV)^3$ and $\la
{\alpha_s \over \pi}
G^2 \ra = 0.012 GeV^4$. The solid, dashed and dash-dotted curves correspond 
to cases (i), (ii) and  (iii), respectively. 
 The curve in case (iii) is the most stable of three cases and 
reproduces the experimental value.
 The others do not show good stability. Also,
 three curves are rather sensitive to the change of $S_N$.


\section{sum rules for the $\pi-N$ coupling constant}
In this section, we look at the two-point correlation function,
\beq
\label{paicorrelation}
\Pi_{\alpha \beta}^{\pi} (q)=i\int d^4x e^{iq \cdot x}
\la0 | T\eta_\alpha (x) \bar{\eta}_\beta (0)
 |\pi^0(p =0) \ra ,
\eeq
where  $|\pi^0 (p=0) \ra$ is a neutral pion state  with vanishing
 four momentum (the soft pion). Since  we are working in the chiral 
limit, 
 this soft pion is simultaneously on shell.

 Since we are taking the matrix element between
 the vacuum and the soft pion in (\ref{paicorrelation}),  only the  
  pseudo-scalar operator $\bar{q} i \gamma_5 q$ survives in OPE 
  for  ${\rm T } \eta(x) \bar{\eta}(0)$.  The pseudo-vector 
operator
 $\bar{q} \gamma_{\mu} \gamma_5 q$ vanishes 
 since $\la 0 | \bar{q} \gamma_{\mu} \gamma_5 q | \pi^0 (p) \ra
 \propto p_{\mu} \rightarrow 0$. Thus
 the relevant terms in 
 OPE read as follows: 
\beq
\label{paiope}
\Pi^\pi(Q^2)&=&\Pi_{dim3}(Q^2)+\Pi_{dim5}(Q^2)+\Pi_{dim7}(Q^2),\\
\label{paiope1}
\Pi_{dim3}(Q^2)&=&-i \gamma_5 {\la0|\bar{d}i \gamma_5 d|\pi\ra 
\over 4 \pi^2}
 \left(1+\frac32{\alpha_s \over \pi}\right)Q^2 \ln{Q^2 \over \mu^2},
 \\
\label{paiope2}
\Pi_{dim5}(Q^2)& =&0,\\
\label{paiope3}
\Pi_{dim7}(Q^2)&=&i \gamma_5 {1 \over 18 Q^2} \la {\alpha_s \over 
\pi} G^2\ra 
\la 0|\bar{d}i \gamma_5 d| \pi \ra .
\eeq
$\mu^2$ dependence of $\alpha_s$ is again implicit in the above 
equations.

The diagrams corresponding to (\ref{paiope1}), (\ref{paiope2})
 and (\ref{paiope3}) are Fig.1(d), 1(e) and 1(f) respectively.
 The Wilson coefficients with $O(\alpha_s)$ corrections
 in (\ref{paiope1}),
(\ref{paiope2})
 and (\ref{paiope3})
  turn out to be identical to 
  (\ref{nucleon.ope1}), (\ref{nucleon.ope2}) and 
(\ref{nucleon.ope3})
 respectively.  This can be  explicitly checked by 
 carrying out OPE with the background field method in the fixed point 
gauge. An alternative way to see this is the plain wave method.
  As an illustration, 
 let's consider dimension 3 operator  (Fig.1(d))
 and expand eq.(\ref{correlation}) 
 in terms of $\bar{d}d$ and  $\bar{d} i \gamma_5 d$;
\beq
\label{dim3ope} 
i \int d^4x e^{i q \cdot x} T(\eta_\alpha (x) \bar{\eta}_\beta (0))
=C_s ({\rm 1})_{\alpha \beta} \bar{d} d+C_{ps} (i \gamma_5)_{\alpha \beta} 
 \bar{d} i \gamma_5 d
 + \cdots
\eeq
where $C_{s,ps}$ is the Wilson coefficient with  $ s (ps)$ denoting 
  scalar (pseudoscalar).
 Note that $\cdots$ are vector, pseudovector and tensor operators.
 Note also that  $\bar{u}u$ and  $\bar{u} i \gamma_5 u$ 
 do not arise for the Ioffe current. We sandwich eq.(\ref{dim3ope})
 by free quark states to extract the Wilson coefficients.
  Applying the projections
 $ {\rm 1}_{\alpha \beta}$ and $ (i \gamma_5)_{\alpha \beta}$ to
 eq.(\ref{dim3ope}), one gets
\beq
& &C_s = {L_1 (q^2 \rightarrow - \infty) 
 \over \la p | \bar{d} d | p \ra},\hspace{1cm} 
L_1=\frac{i}4 \int  d^4x e^{iq\cdot x}
\la p| T(\eta(x) \bar{\eta}(0)) |p \ra \\
& &C_{ps}={L_{i\gamma_5}(q^2 \rightarrow - \infty )
  \over \la p | \bar{d}i \gamma_5 d |p 
\ra},  \hspace{1cm}
 L_{i \gamma_5}=\frac{-i}4 \int d^4x e^{i q\cdot x}
 \la p| T(\eta(x) i \gamma_5 \bar{\eta}(0))| p \ra
\eeq
 with $|p\ra$ being a quark state with momentum $p$.
 Since massless QCD does not flip chirality
 in perturbation theory,
 $\la p | \bar{d} d | p \ra \sim \la p | \bar{d}i \gamma_5 d |p 
\ra $ and $L_1 \sim L_{i \gamma_5}$ except for trivial kinematical factors.
 Thus $C_s = C_{ps}$ is obtained even when $\alpha_s$ corrections
 are included.

  $\Pi^{\pi}(Q^2)$ satisfies the  dispersion relation
\beq
\label{pai.dispersion}
\Re \Pi^{\pi}(Q^2)&=&{1 \over \pi} \int {\Im\Pi^{\pi}(s) \over 
s+Q^2} ds +   
{\rm subtraction}.
\eeq
The hadronic imaginary part $\Im \Pi^{\pi}(s)$ has a nucleon pole 
 and the 
 continuum. The pole part is parametrized by assuming
 the $\pi-N-N$ vertex  ${\cal L}_{int} = g_{\pi N} \bar{N} i 
\gamma_5
 \vec{\tau}\cdot \vec{\pi} N $ as was done in \cite{yazaki,RRY}.
 The continuum part is extracted from the analytic continuation
 of OPE. In total,
\beq
\label{pionimagi}
\Im \Pi^{\pi}(s)&=&i \gamma_5 \pi \lambda^2_N g_{\pi N} 
\delta(s-M_N^2)
 \nonumber \\
& & - i\gamma_5 \pi {s \over 4 \pi^2}
\left(1+\frac32 {\alpha_s \over \pi }\right) \theta(s-S_\pi)
\la 0 |\bar {d} i\gamma_5 d|\pi^0 \ra ,
\eeq
where $S_\pi$ is the continuum threshold.
 Note that $S_\pi$ does not have to be the same with
  $S_N$.

 Putting Eq.(\ref{pionimagi}) and Eq.(\ref{paiope})
 into (\ref{pai.dispersion}) and making the Borel improvement,
 one obtains  the following sum rule:
\beq
\label{g-coup}
g_{\pi N}={e^{M_N^2 \over M^2} \over \lambda^2_N}
        \left\{{M^4 \over 4 \pi^2} 
E_1(\frac{S_\pi}{M^2})\left(1+\frac32 
{\alpha_s \over \pi}\right) - \frac1{18}\la{\alpha_s \over \pi} G^2
  \ra \right \}
 \left(-\frac1{f_\pi} \la \bar{d} d \ra \right),
\eeq
where $f_\pi $ is the pion decay constant (93 MeV), and
$\la 0|\bar{d}i\gamma_5 d |\pi^0\ra$ has been rewritten by  the
 soft pion theorem 
\beq
\la 0|\bar{d}i \gamma_5 d| \pi^0(p=0) \ra =\frac1{f_\pi}\la 
\bar{d}d \ra.
\eeq
One can get rid of the coefficient 
 ${e^{M_N^2 /M^2} /\lambda^2_N}$ in (\ref{g-coup}) by using
 the nucleon sum rules (\ref{nucleonQSR1}) or (\ref{nucleonQSR2}).  
Thus
 one arrives at two different sum rules for $g_{\pi N}$:
\beq
\label{evensum}
g_{\pi N}^{even}
={\pi^2 \left\{M^4 \left(1+\frac32 {\alpha_s(M^2)
 \over \pi} \right)E_1({S_\pi\over M^2})
 -{b \over 18}\right\} \left(-\frac1{f_\pi}\la \bar{d}d \ra 
\right)
\over 
{M^6 \over 8}E_2({S_N \over M^2})
 \left(1+(\frac{53}{12}+\gamma_E){\alpha_s(M^2) 
\over \pi} \right) + 
{b M^2 \over 32}E_0({S_N \over M^2})  + {a_u^2 \over 6} 
\left(1-(\frac56-\frac13 \gamma_E){\alpha_s(M^2) \over  \pi} 
 \right )} ,\nonumber \\
\eeq
which is obtained from the ``even" sum rule (\ref{nucleonQSR1}) 
 for the nucleon,
 and 
 \beq
\label{oddsum}
g_{\pi N}^{odd}={M_N \left 
\{E_1({S_\pi \over M^2})\left ( 1+\frac32 {\alpha_s (M^2) \over 
\pi} \right)
 - {b \over 18 M^4} \right \}
 \over 
f_\pi 
 \left \{E_1({S_N\over M^2})\left ( 1+\frac32 {\alpha_s(M^2) \over 
\pi} \right)
 - {b \over 18 M^4} \right \} },
\eeq
which is obtained from the ``odd" sum rule (\ref{nucleonQSR2})
 for the nucleon. 
 For $M_N$ in eq.(\ref{oddsum}), we  just take the experimental number
 instead of reexpressing $M_N$ by the Borel mass $M$
 through the nucleon sum rule.  Even if one uses $M_N(M)$
 in eq.(\ref{oddsum}), the
  Borel curve for $g_{\pi N}^{odd}$ is not affected so  much
  as far as one adopts 
the odd nucleon sum rule (case (iii) in section II).
 Note here that, 
if one assumes $S_\pi=S_N $, Eq.(\ref{oddsum}) gives 
 $g_{\pi N}^{odd}=M_N/f_{\pi}$, namely the GT relation
 with $g_A=1$.
     
It is in order here to remark some difference  of the present
  work from that of ref. \cite{yazaki,RRY}.
 First of all, we have carried out  OPE up to dimension 7
 both for  $g_{\pi N}$ and the nucleon mass, while 
 only the lowest dimension 
operator, namely $ \bar{q}i \gamma_5 q $, is taken into account
 for the $g_{\pi N}$ sum rules in ref.\cite{yazaki,RRY}.
 Secondly, $\alpha_s$ corrections to the Wilson coefficients
  are not taken into account in
 \cite{yazaki,RRY}. Thirdly,
 the continuum thresholds $S_{\pi}$ and $S_N$ are completely
 neglected in \cite{yazaki,RRY} for the $g_{\pi N}$ sum rules.


\section{Borel analysis for the $\pi -N$ coupling constant}

\subsection{ The determination of the thresholds $S_\pi$ and  
$S_N$}

As we have mentioned in section II, it is difficult to 
  determine $S_N$ from the Borel sum rules of the nucleon.
  In this section, we will utilize the finite energy sum rules
 (FESR)  to give a constraint on $S_N$ as well as $S_{\pi}$.

The FESR has  a general form \cite{KPT} ,
\beq
\label{FESR}
\int_0^{S_0} ds \ s^n \ \Im \Pi_{OPE}(s)=\int^{S_0}_0 ds \ s^n \ 
\Im \Pi_{phen.}(s),
\eeq
where $\Im \Pi_{OPE}(s) $ is an imaginary part
 obtained by the analytic continuation of $\Pi$ in OPE, 
$\Im \Pi_{phen.}(s)$ is
the phenomenological 
 imaginary part and $S_0$ is the continuum threshold (either
 $S_N$ or $S_{\pi})$.
  For the nucleon, by using $\Pi_1$ with $n=0,1$ and
 $\Pi_2$ with $n=0$, one gets
 \cite{ovchinnikov}
\beq
\label{NFESR}
& & 64 \pi^4 \lambda^2_N= \left (1 + \frac{25}4 {\alpha_s \over 
\pi}
\right ){S_N^3 \over 3}
+ 2 \pi^2 \la {\alpha_s \over \pi} G^2 \ra  S_N 
+ {128 \over 3} \pi^4 \la \bar{q} q \ra^2 \left (1-\frac56 {\alpha_s 
\over \pi}
 \right) , \\
& &
64 \pi^4 \lambda^2_N M_N
= -8 \pi^2 \la \bar{q}q \ra 
\left (1+ \frac32 {\alpha_s \over \pi} \right) S_\pi^2
 +{32 \over 9} \pi^4 \la \bar{q}q\ra \la {\alpha_s \over \pi} G^2 \ra ,\\
\label{NFESRB}
& &
64 \pi^4 \lambda^2_N  M_N^2
= \left (1+\frac{37}6 {\alpha_s \over \pi}\right ){S_N^4 \over 4}
 +\pi^2 \la {\alpha_s \over \pi} G^2 \ra S_N^2
 -{128 \over 9} \pi^4  \la \bar{q}q\ra^2  {\alpha_s  \over \pi} S_N .
\eeq 
  The renormalization point here is chosen as  $\mu^2=S_N$.

 By using the standard values of the condensates
 $\la \bar{q}q\ra (1{\rm GeV}^2)  =-(225\pm 25MeV)^3$ and
 $\la {\alpha_s \over \pi}G^2 \ra =0.012GeV^4$, 
 we solved  Eq.(\ref{NFESR}) $\sim$ Eq.(\ref{NFESRB}) numerically.
  Table 1 shows  the results for four different values
 of $\la \bar{q} q \ra$.

\vspace{0.8cm}

\begin{center}

\begin{tabular}{|c|c|c|c|c|} \hline
$\la \bar{q}q \ra$ & $(-0.250GeV)^3$  & $(-0.240GeV)^3$  &
 $(-0.225GeV)^3$ & $(-0.200GeV)^3$  \\ \hline
$S_N(GeV^2)$ & 1.77 & 1.60 & 1.34 & 0.887  \\ \hline
$\lambda_N(GeV^3)$ & 0.0267 &0.0235 & 0.0187 & 0.0118 \\ \hline
$M_N (GeV) $ & 0.997 & 0.940 &  0.845  & 0.645\\ \hline
\end{tabular}\\

\vspace{0.5cm}

Table1: $S_N, \lambda_N, M_N $ obtained from Eq.(\ref{NFESR}) $\sim$
 (\ref{NFESRB}) with four different values of $\la \bar{q}q\ra$.
$\la \bar{q}q\ra=-(0.2402GeV)^3$ reproduces the nucleon mass.

\end{center}

\vspace{0.8cm}

 \ From this table, we choose  $S_N = 1.34 - 1.77$
 as physical range where the nucleon mass is reasonable reproduced 
  within $\pm 10\%$ errors.
 Our $S_N$ is  smaller than that usually used in
 the literatures \cite{henley,Ioffe,YHHK,KPT}.
  However, the $\alpha_s$ corrections
are not 
 taken into account in these references.
  The effect of the  $\alpha_s$ correction
 to the spectral parameters can be
 explicitly seen by expanding the solutions of Eq.(\ref{NFESR}) 
 up to linear in
 $\alpha_s$;
\beq
\label{a1}
& &\lambda^2_N
=\lambda_0^2 \left(1-26.2 (-16\pi^2 \la \bar{q}q \ra)^{-\frac43}
 \la {\alpha_s \over \pi} G^2 \ra \right )(1-3.08{\alpha_s \over \pi}), \\
\label{a2}
& &S_N=S_N^0 \left(1-21.2 (-16 \pi^2 \la \bar{q} q \ra)^{-\frac43}
 \la {\alpha_s \over \pi} G^2 \ra \right )(1-3.26{\alpha_s \over 
\pi}), \\
\label{a3}
& &M_N=M_N^0 \left(1-18.6 (-16 \pi^2 \la \bar{q}q \ra)^{-\frac43}
 \la {\alpha_s \over \pi} G^2 \ra \right )(1-1.94{\alpha_s \over 
\pi}), 
\eeq
where $\lambda_0^2 = 4\la \bar{q}q\ra^2$, $S_N^0
=(640 \pi^4 \la \bar{q}q \ra^2)^\frac13$ and  
$M_N^0=(-\frac{25}{2} \pi^2 \la \bar{q} q \ra)^\frac13$,
 which are the solutions when the $\alpha_s$ corrections and
the gluon condensate 
 are neglected.
(\ref{a1})-(\ref{a3}) show that the $\alpha_s $ corrections
 tend to  reduce the observables by considerable amount
 particularly in $S_N$.

 Next, we estimate  $S_\pi$ by taking the $n=0$ FESR of 
$\Pi^{\pi}$;
\beq
\label{PAIFESR}
\lambda_N^2 g_{\pi N}=\left \{ {1 \over 8 \pi^2} S_\pi^2 \left(1 + 
\frac32 
{\alpha_s(S_\pi) \over \pi} \right)
- \frac1{18} \la {\alpha_s \over \pi} G^2\ra \right \}
\left (-\frac1{f_\pi} \la\bar{d}d \ra \right) .   
\eeq
Since the FESR is rather sensitive to the 
 structure of the continuum compared to the Borel sum rule and we
do not know much about the detailed structure
 of the continuum for (\ref{paicorrelation}),
 we just limit ourselves to the $n=0$ sum rule (local duality 
relation)
 for safety.
 To roughly evaluate the range of $S_{\pi}$, we simply
 put $g_{\pi N}=13.4$ 
 in (\ref{PAIFESR}) with $\lambda_N$
being determined
 in the nucleon FESR.
 The result is given in Table 2 for three different values of the 
condensate:

\vspace{0.8cm}

\begin{center}

\begin{tabular}{|c|c|c|c|} \hline
 &$\la \bar{q}q \ra(GeV^3)$ & $S_N(GeV^2)$ & $S_\pi(GeV^2)$  \\ 
\hline
set 1 & $-(0.250)^3 $  & 1.77 & 1.98  \\ \hline
set 2 & $-(0.240)^3 $ & 1.60 & 1.85 \\ \hline
set 3 & $-(0.225)^3 $ & 1.34 & 1.62\\ \hline
\end{tabular}\\

\vspace{0.5cm}

Table 2: $S_N, S_\pi$ with three different values of $\la \bar{q} q \ra$.
$S_\pi$ is obtained by substituting $g_\pi$=13.4 into
  (\ref{PAIFESR}).   

\end{center}
\vspace{0.8cm}

 From Table 2, one finds that $S_{\pi} $ is
always greater than $S_{N}$.
 This is consistent with the Borel sum rule $g_{\pi N}^{odd}$
 in (\ref{oddsum}) which tells us that $g_{\pi N} > M_N/f_{\pi}$
 only when $S_{\pi} > S_{N}$. In  subsections below, we will
 examine the Borel stability of 
  $g_{\pi N}$ with the parameter sets obtained
 in Table 2.

\subsection{ Borel analysis for $g_{\pi N}^{even}$ }
  In Fig.3,  $g_{\pi N}^{even}$ is shown as a function
 of $M^2$.
 The solid, dashed and  dash-dotted curves correspond to
  set 1, set 2 and set 3 in Table 2, respectively.  
 $g_{\pi N}^{even}$ in  Fig.3(a) includes the $\alpha_s$
 corrections to the Wilson coefficients,
 while they are neglected  in Fig.3(b)
  except for the gluon condensate.

 $g_{\pi N}^{even}$ 
 has a sizable $M^2$ variation and good Borel stability is not seen in
 Fig.3. 
 By comparing Fig.3(a) with Fig.3(b), one finds that 
 the $\alpha_s$ corrections
 improve  the Borel stability only slightly.
 The effect of the higher
dimensional operator to the Borel curve is also small.
 In fact,
 ${2 \pi^2 \over 9 M^4} \la {\alpha_s \over \pi} G^2\ra 
/ \{E_1({S_\pi \over M^2}) (1+{3 \over 2}{\alpha_s \over \pi})\}$,
 which is a  ratio of the dimension 3  term  and the dimension 7 term 
in eq.(\ref{g-coup}), is  about 4 \% at $M^2 \sim 1 {\rm GeV}^2$.
 (Note that dimension 5 terms do not arise for the Ioffe current.)

 Although $g_{\pi N}^{even}$ in eq.(\ref{evensum})  is proportional to 
 $\la \bar{q}q\ra$,
 three curves in
Fig.3, which correspond to different values of  $\la \bar{q}q\ra$,
  almost
overlap with each other. This is because
 the change of $\la \bar{q}q\ra$ is  compensated
 by the changes of $S_{\pi ,N}$.


 In Fig.4,  $g_{\pi N}^{even}$ for set 2 is shown 
 with $S_{\pi}$ being changed by $\pm 10 \%$.
 The solid, dashed and dash-dotted curves correspond
 to $S_\pi=1.85 \times 1.1, 1.85$ and $ 1.85 \times 0.9$, respectively.
 $g_{\pi N}^{even}$ increases as $S_{\pi}$ increases, which
 is consistent with the prediction of FESR in
 eq.(\ref{PAIFESR}).
 
\subsection{Borel analysis for $g_{\pi N}^{odd}$}
 Fig.5(a),(b) show  $g_{\pi N}^{odd}$ as a function of $M^2$.
 The solid, dashed, dash-dotted curves 
correspond to set 1, set 2, set 3, respectively.
 $g_{\pi N}^{odd}$ in  Fig.5(a) includes $\alpha_s$
 corrections to the Wilson coefficients, while they 
 are neglected in Fig.5(b)
  except for the gluon condensate.

$g_{\pi N}^{odd}$ has apparently  better Borel stability
  than $g_{\pi N}^{even}$, 
 but still sizable $M^2$ variation is seen.
 The  $\alpha_s$ corrections do  not affect the Borel stability
 much, since the same $\alpha_s$ correction appears both
 in numerator and denominator in eq.(\ref{oddsum}).

$g_{\pi N}^{odd}$ is rather sensitive to the
 change of the parameter sets , in particular the 
 $S_{\pi}/S_N$.
 To see the 
 effect of $S_{\pi,N}$ on $g_{\pi N}^{odd}$  in more detail, 
 we expand
 eq.(\ref{oddsum})  up to {\it O}$({1 \over M^2})$:
\beq
g_{\pi N}^{odd}& \simeq &{M_N \over f_\pi}{(S_\pi^2-\frac{b}9)
 \over (S_N^2-\frac{b}9)}
 \left \{1-{2 \over 3 M^2} \left({S_\pi^3 \over S_\pi^2-\frac{b}9 }-
{S_N^3 \over S_N^2-\frac{b}9} \right) \right \}\nonumber  \\
\label{oddsuma}
 & \simeq & {M_N \over f_\pi}({S_\pi \over S_N})^2 
\left \{1-{2 \over 3M^2}(S_\pi -S_N) \right \} \ \ ,
\eeq
where we have neglected  small $\alpha_s$ corrections
 and used the fact $S_{\pi N}^2 \gg b/9$.
 The approximate formula eq.(\ref{oddsuma})
 is in  good agreement with the exact one eq.(\ref{oddsum}) 
 in 10 \% for $M^2 > 1.0 {\rm GeV}^2$.
 
 When $S_\pi > S_N$, $M^2$ independent term in (\ref{oddsuma})
 gives $g_{\pi N}^{odd} = (S_{\pi}/S_N)^2 (M_N/f_{\pi})$
 which is larger than
 $ M_N/f_{\pi}$. The leading $1/M^2$ correction
  reduces $g_{\pi N}^{odd}$ slightly.
 The experimental value for $g_{\pi N}=13.4$ is obtained
 when $M^2 = 1.6{\rm GeV}^2$ for the parameter set 3.

 In Fig.6,  $g_{\pi N}^{odd}$ for set 2 is shown 
 with $S_{\pi}$ being changed by $\pm 10 \%$.
 The solid, dashed and dash-dotted curves correspond
 to $S_\pi=1.85 \times 1.1, 1.85$ and $ 1.85 \times 0.9$, respectively.
 $g_{\pi N}^{odd}$ increases as $S_{\pi}$ increases, which
 is consistent with the prediction of  FESR in
 eq.(\ref{PAIFESR}) and also with the approximate formula (\ref{oddsuma}).

\subsection{Comparison of $g_{\pi N}^{even}$ and $g_{\pi N}^{odd}$}

As we have already mentioned, the Borel stability
 for $g_{\pi N}^{odd}$ is better than 
 $g_{\pi N}^{even}$.  This is consistent with the fact that
 the Borel curve  for  $M_N$ is
 most stable in ``odd" sum rule (case (iii) in Fig.2).
 Also the absolute value of $g_{\pi N}^{odd}$ 
 is larger than $g_{\pi N}^{even}$ for
 appropriate range of $M^2$: e.g. $g_{\pi N}^{odd}=$11.3 $-$
12.8 versus $g_{\pi N}^{even}=$9.02 $ -$ 9.10 at $M^2=1 {\rm GeV}^2$.

In previous subsections, $S_{\pi} = 1.85 GeV^2$ has been used
 as a standard value to study the Borel stability.
 Alternative way to make the Borel analysis is 
 to eliminate $S_{\pi}$ from eq.(\ref{evensum})
 and eq.(\ref{oddsum}) using 
 eq.(\ref{PAIFESR}), and then to solve $g_{\pi N}$
 self-consistently for each $M^2$.  By this procedure,
  we found that there exists no solution 
 satisfying eq.(\ref{evensum}) and eq.(\ref{PAIFESR}) simultaneously,
 while there exists  solutions of 
  eq.(\ref{oddsum}) and eq.(\ref{PAIFESR}) which are given in
 Fig.7. This result again confirms that
 $g_{\pi N}^{odd}$ is better starting point
 to study the $\pi-N$ coupling constant than $g_{\pi N}^{even}$.
   
\section{conclusion}
In this article we have made extensive Borel and FESR analyses 
 of $ g_{\pi N}$ by taking into account  higher dimensional operators, 
  $\alpha_s$ corrections and the continuum threshold.
 None of them has been considered in the previous
 analyses which led to $g_{\pi N}=M_N / f_\pi$ \cite{yazaki,RRY}.

What we have found are summarized as follows:\\
(a) The higher dimensional operators up to dim. 7
  and the $\alpha_s$ corrections 
play no crucial role for the Borel stability of  $g_{\pi N}$. \\
(b)  $g_{\pi N}^{odd}$ is more appropriate for examining $g_{\pi N}$  than
$g_{\pi N}^{even}$, since the former has better Borel stability.
 This fact is also consistent with the fact that ``odd" sum rule
 for$M_N$ has a best stability.\\
(c) $g_{\pi N}$ is most sensitive to the
 ratio $S_{\pi}/S_N$, and both the FESR and Borel sum rules
 tell us that $S_{\pi}/S_N > 1$ 
 is a crucial ingredient to reproduce
 the experimental
 $g_{\pi N}$.\footnote{This point
 may have some relation to the Adler-Weisberger sum rule \cite{current}
 which tells us that $g_A > 1$ (or equivalently
 $g_{\pi N} > M_N/f_{\pi}$) is obtained only when
 the continuum contribution in the $\pi-N$ channel is taken into account.}

We also found that  
 the Borel stability of $g_{\pi N}^{odd}$, even if dim. 7 operator is
 taken into account,  is not
satisfactory enough
 to determine the pion-nucleon coupling constant precisely.
 For the nucleon sum rule,  there have been
 some attempts to improve the Borel stability 
 such as the modification of the Ioffe current \cite{Dosch}
 and the inclusion of the instantons \cite{Forkel}.
 In particular,  instantons 
  improve the
 nucleon Borel sum rules considerably at low $M^2$ region,
 so it will be an interesting problem to study
  $g_{\pi N}$ with instanton contribution in the future.

\acknowledgments

This work was supported in part by the Grants-in-Aid of the Ministry of 
Education (No. 06102004).  T. H. thanks Institute for Nuclear Theory
 at the University of Washington for its hospitality
 and the Department of Energy for partial support
 during the completion of this work.

\newpage

\appendix
\setcounter{equation}{0}
\renewcommand{\theequation}{A.\arabic{equation}}
\section{}
The Borel transform is defined as,
\beq
\hat{B}= { (-1)^n (Q^2)^n \over (n-1)!} \left ({d \over dQ^2} 
\right)^n,\ \ \ \    Q^2=-q^2 \ \ \ ,
\eeq
with $Q^2 \rightarrow \infty, n \rightarrow  \infty,$ and $ {Q^2 \over 
n}= M^2$ being fixed.

When $\hat{B}$ is applied to the correlation function 
$ \Pi (Q^2)={1 \over \pi}\int   { {\rm Im}\Pi(s) \over s+Q^2} $, it
 leads to 
 \beq  
  \hat{B}\Pi(Q^2)={1 \over \pi M^2} \int \ ds \  \Im\Pi (s) e^{-{s \over 
M^2}}
\eeq
This shows that the Borel transform tends to suppress the high energy
 contribution.

Some useful formula are
\beq
& & \ \ \hat{B}\left( 1 \over Q^2 \right)^k={1 \over (k-1)!} \left(1 
\over M^2 \right)^k ,\\
& & \ \  \hat{B}(Q^2)^k \log Q^2=(-1)^{k+1}\Gamma (k+1) (M^2)^k ,\\
& & \ \  \hat{B}{1 \over s +Q^2} = {1 \over M^2} e^{-{s \over M^2}},\\
& & \ \  \hat{B}{\log Q^2 \over Q^2} = {1 \over M^2} (\log M^2 
-\gamma_E), \\
& & \ \  \hat{B} (\log Q^2)^2=-2 \log M^2 + 2 \gamma_E ,\\ 
& & \ \  \hat{B} Q^2(\log Q^2)^2 =2 M^2( \log M^2 - \gamma_E +1), \\
& &  \ \ \hat{B} (Q^2)^2 (\log Q^2)^2=(M^2)^2(-4 \log M^2 + 4 
\gamma_E -6). 
\eeq
\\
%
\newpage
\centerline{Figure Captions} 
\noindent
{\bf Fig.1}\\
 OPE  up to dimension 7 operators for the correlation of Ioffe current.
 Wavy lines denote gluon lines, broken lines denote
 the quark/gluon condensate.\\
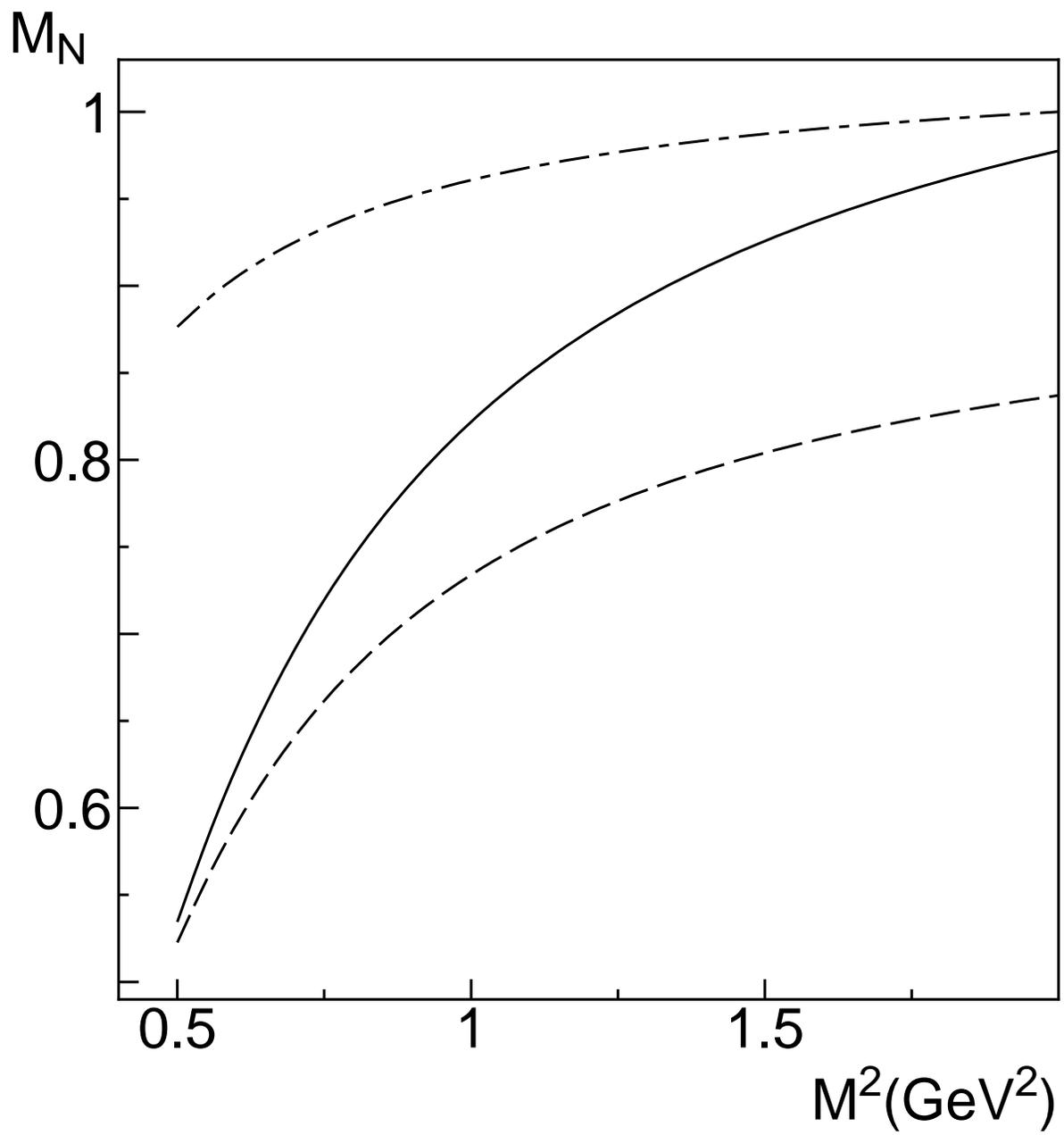
{\bf Fig.2}\\
 $ M_N$ (nucleon mass) as a function of the Borel mass squared $M^2$.
 The solid, dashed,  dash-dotted lines correspond
 to the cases (i), (ii) and (iii), respectively. $\la \bar{q} q \ra
=-(0.240GeV)^3 $, $\la {\alpha_s \over \pi} G^2 \ra=0.012GeV^4$ 
 and $S_N = 1.60 GeV^2$ are used.\\
{\bf Fig.3(a),(b)}\\
$g_{\pi N}^{even}$ as a function of  $M^2$. 
The solid, dashed and dash-dotted lines correspond to
 set 1, set 2 and set 3, respectively.
 $\alpha_s$ corrections are taken into account in Fig.3(a), while
 they are neglected in Fig.3(b) except for gluon condensate.\\
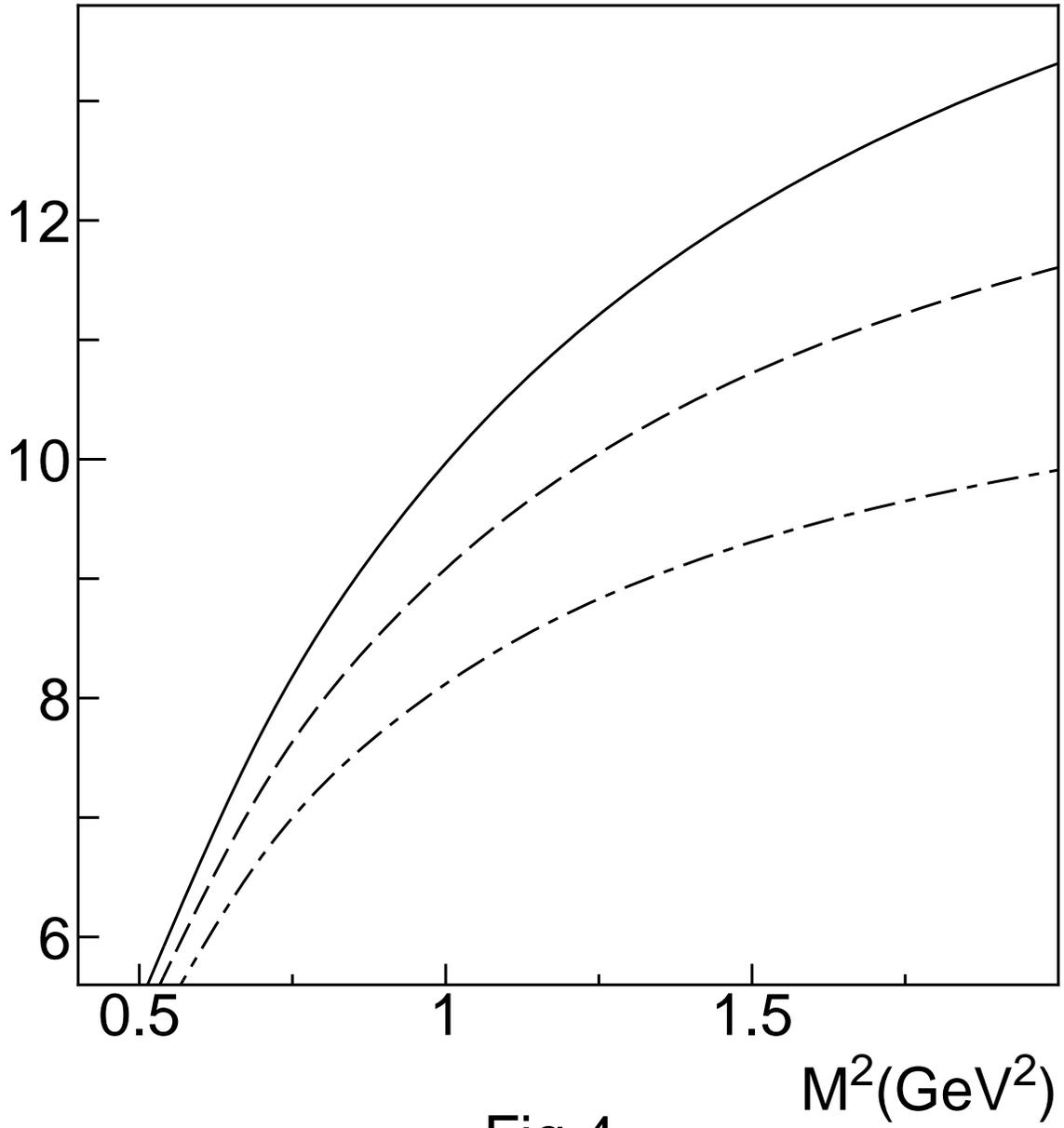
{\bf Fig.4}\\
$g_{\pi N}^{even}$ 
 with $\la \bar{q} q\ra=-(0.240GeV)^3$
as a function of  $M^2$.
The solid, dashed and dash-dotted lines correspond
 to $S_\pi =1.85 \times 1.1, 1.85$ and $ 1.85 \times 0.9$, respectively.\\
{\bf Fig.5(a),(b)}\\
$g_{\pi N}^{odd}$ as a function of  $M^2$. 
The solid, dashed and dash-dotted lines correspond to
 set 1, set 2 and set 3, respectively.
 $\alpha_s$ corrections are taken into account in Fig.5(a), while
 they are neglected in Fig.5(b) except for gluon condensate.\\
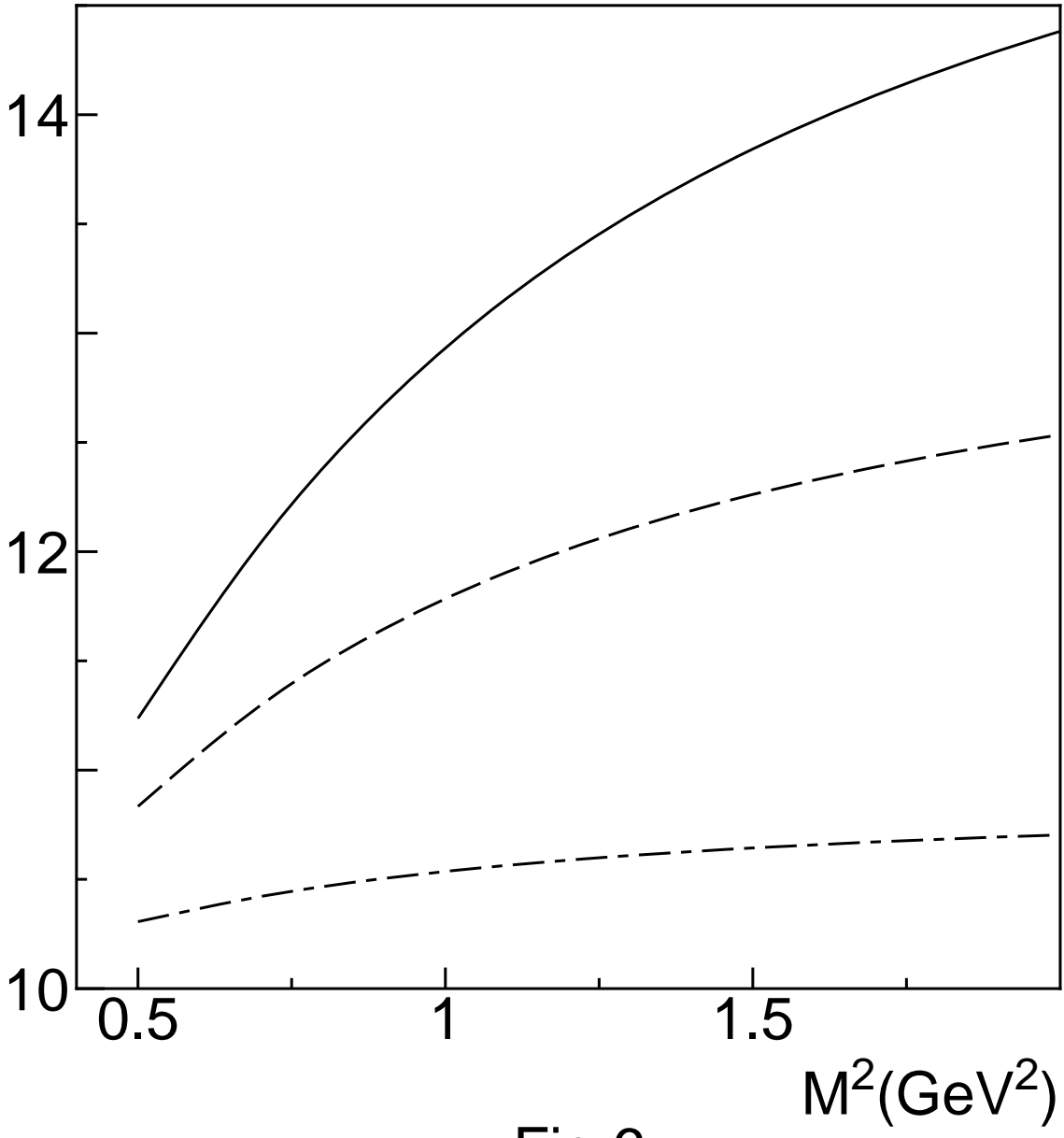
{\bf Fig.6}\\
$g_{\pi N}^{odd}$ 
 with $\la \bar{q} q\ra=-(0.240GeV)^3$
as a function of  $M^2$.
The solid, dashed and dash-dotted lines correspond
 to $S_\pi =1.85 \times 1.1, 1.85$ and $ 1.85 \times 0.9$, respectively.\\
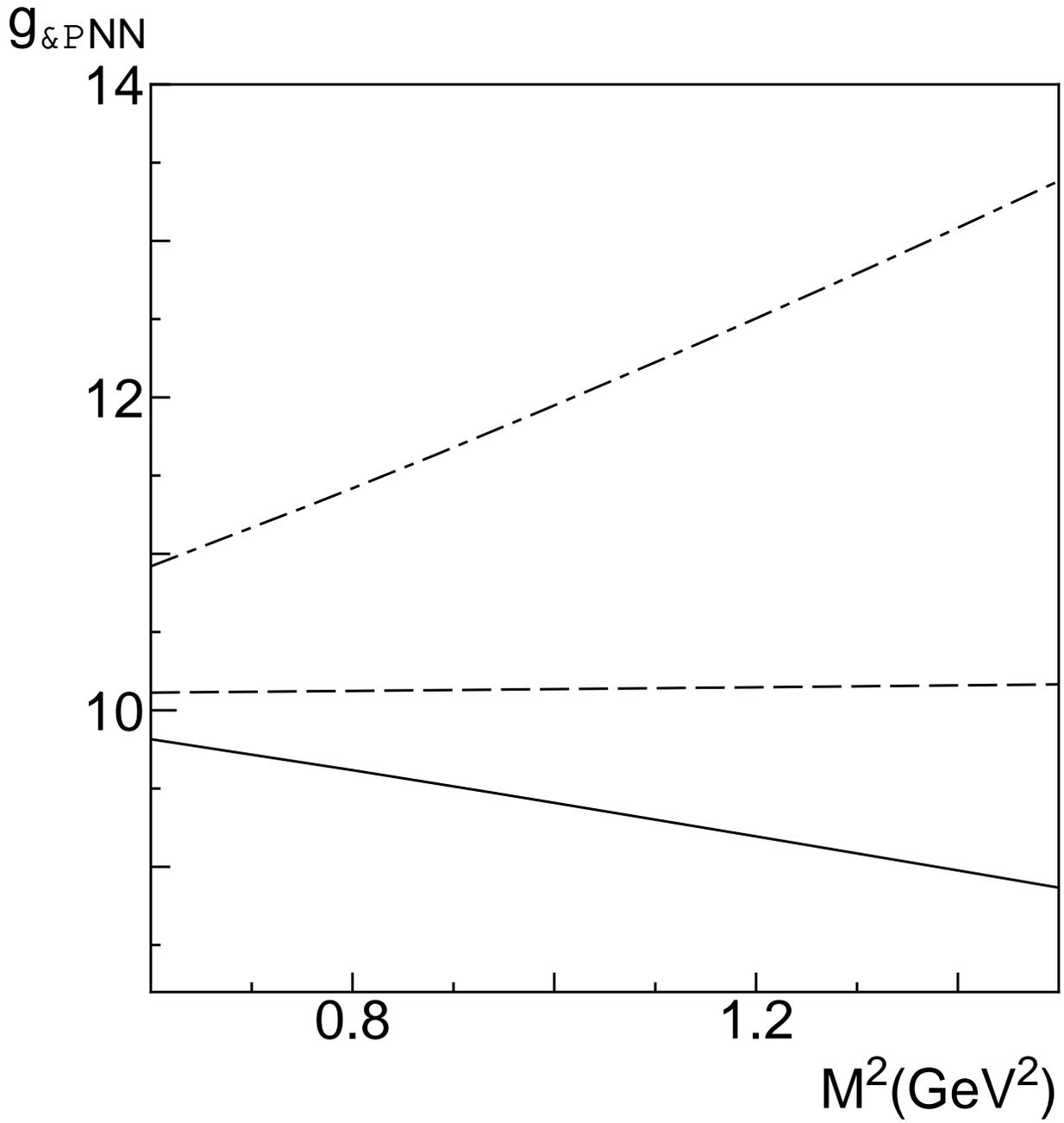
{\bf Fig.7}\\
$g_{\pi N}$ as a function  of  $M^2$.
$S_\pi$ in eq.(\ref{PAIFESR}) with $\lambda_N$ being determined in the nucleon
FESR is used for $S_\pi$ in eq.(\ref{oddsum}).
The solid, dashed and dash-dotted lines correspond to $g_{\pi N}^{odd}$ with 
$\la \bar{q}q \ra=-(0.25GeV)^3,
 -(0.240GeV)^3$ and $ -(0.225GeV)^3$, respectively.\\
%

\newpage

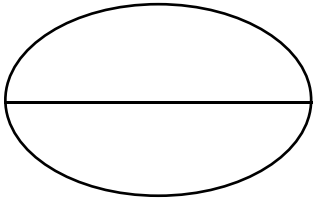
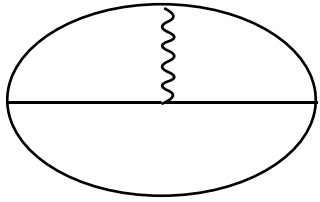
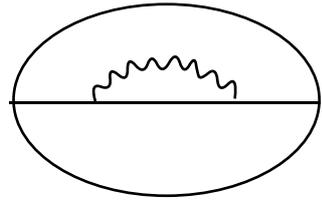
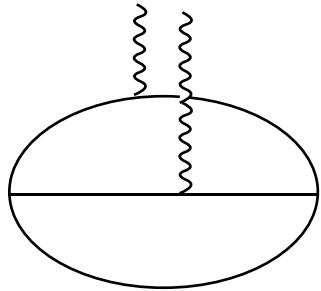
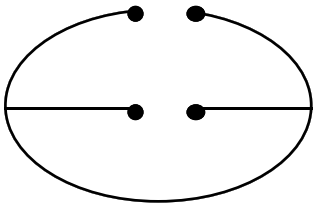
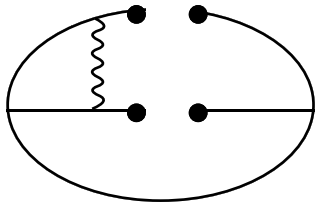
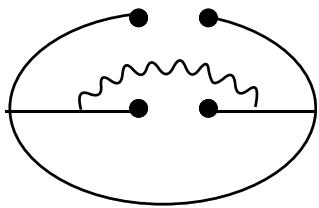
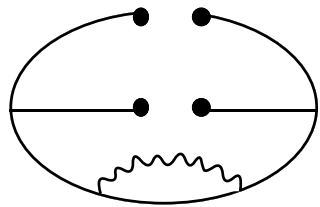
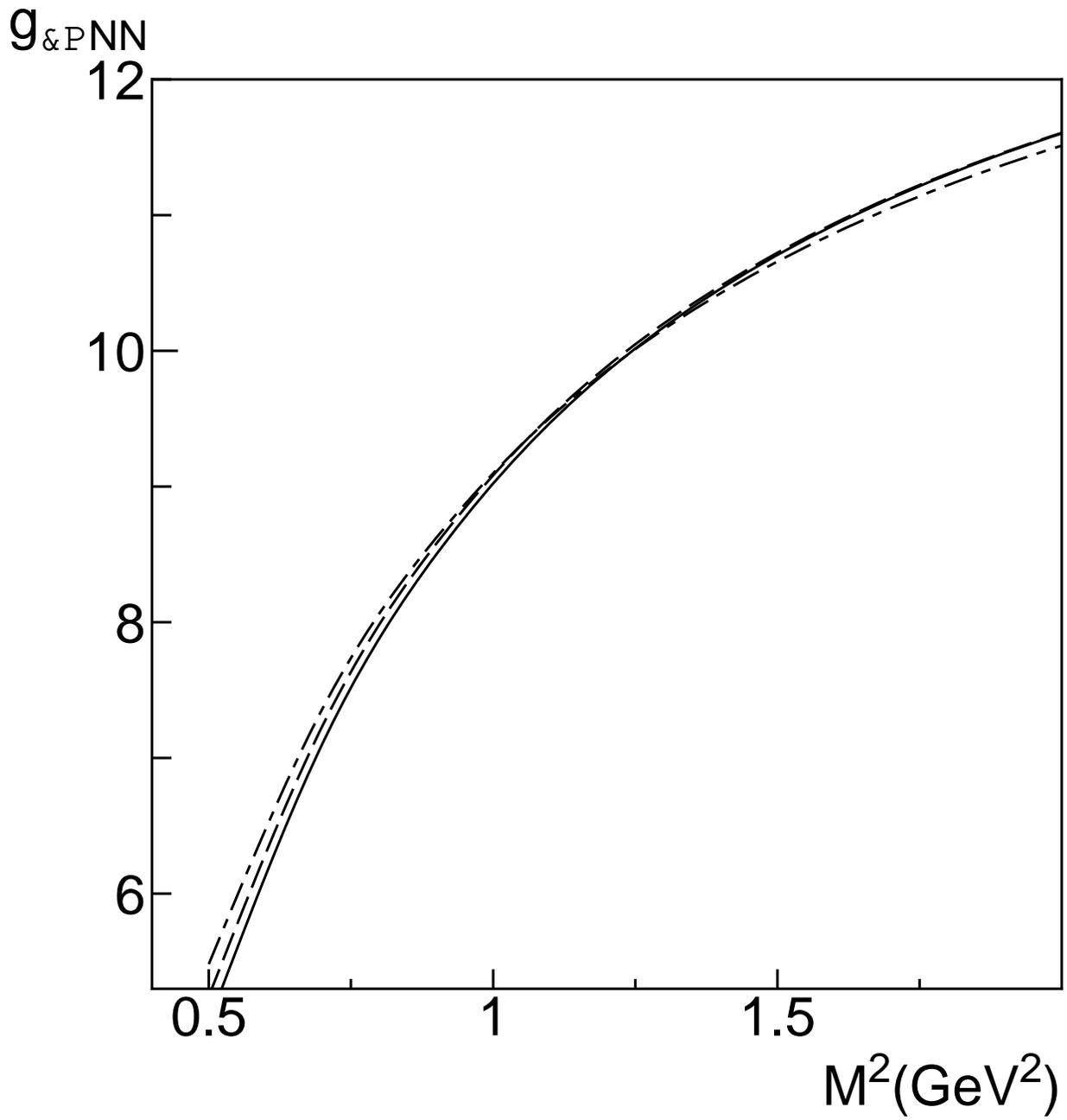
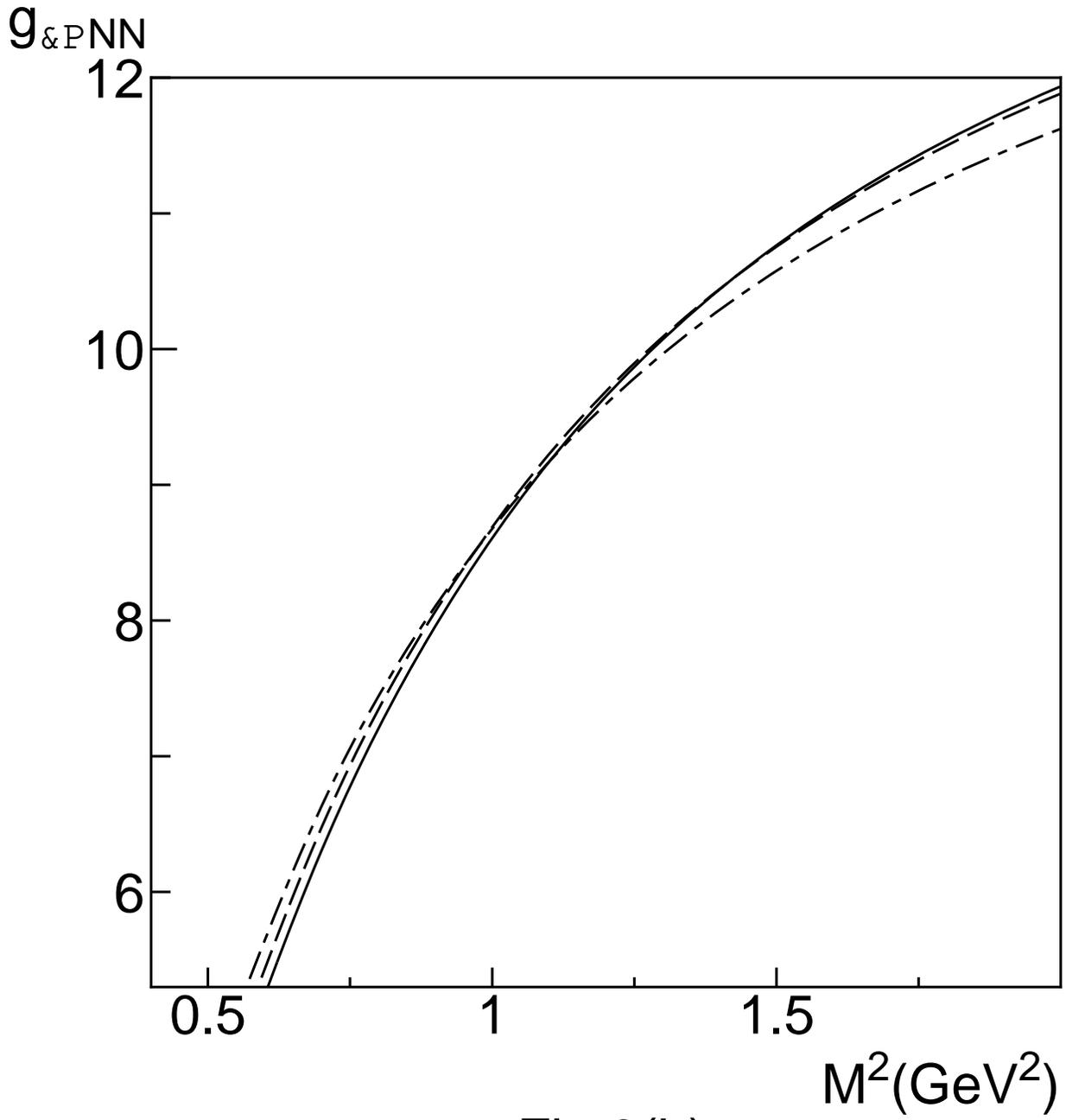
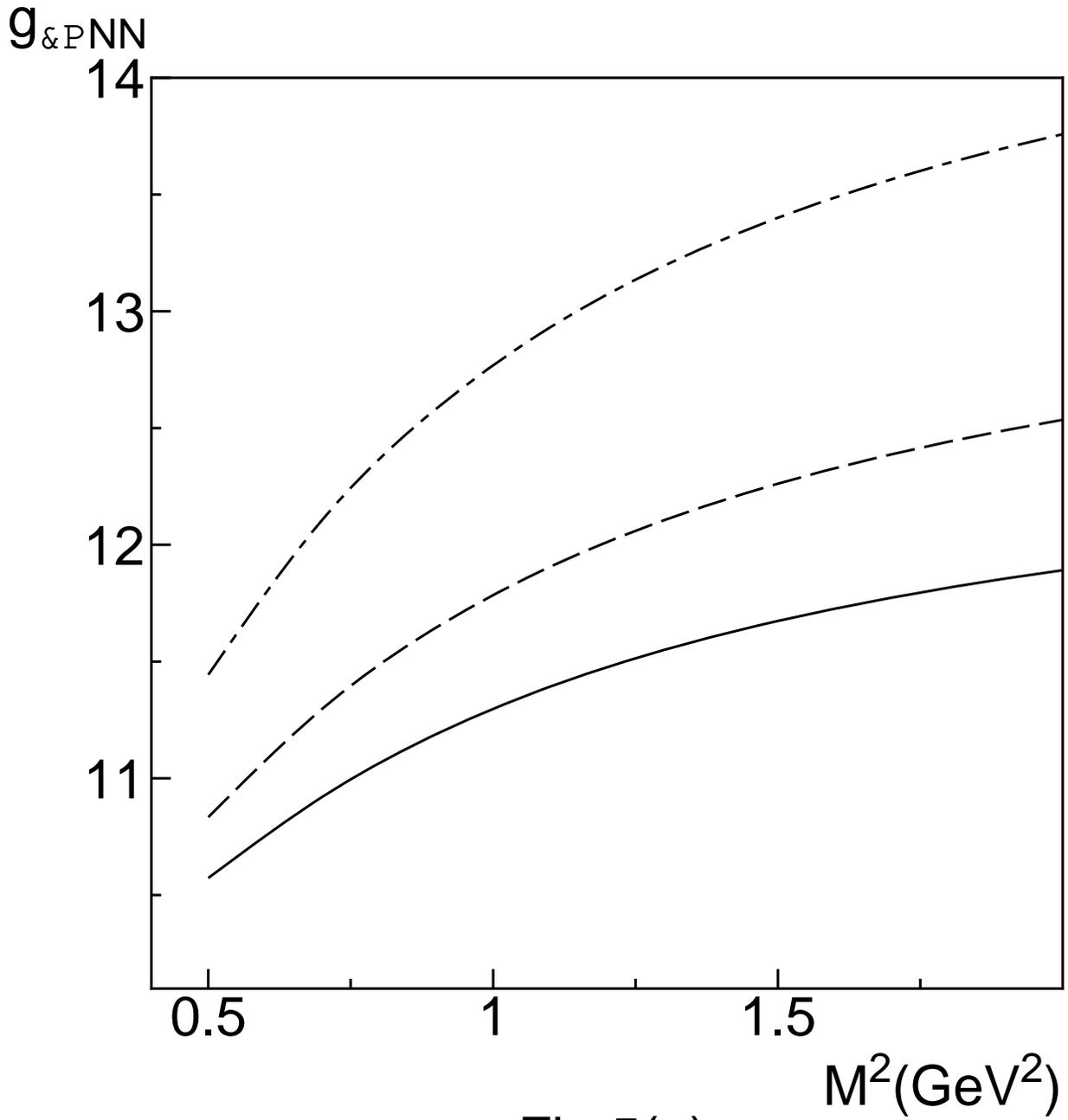
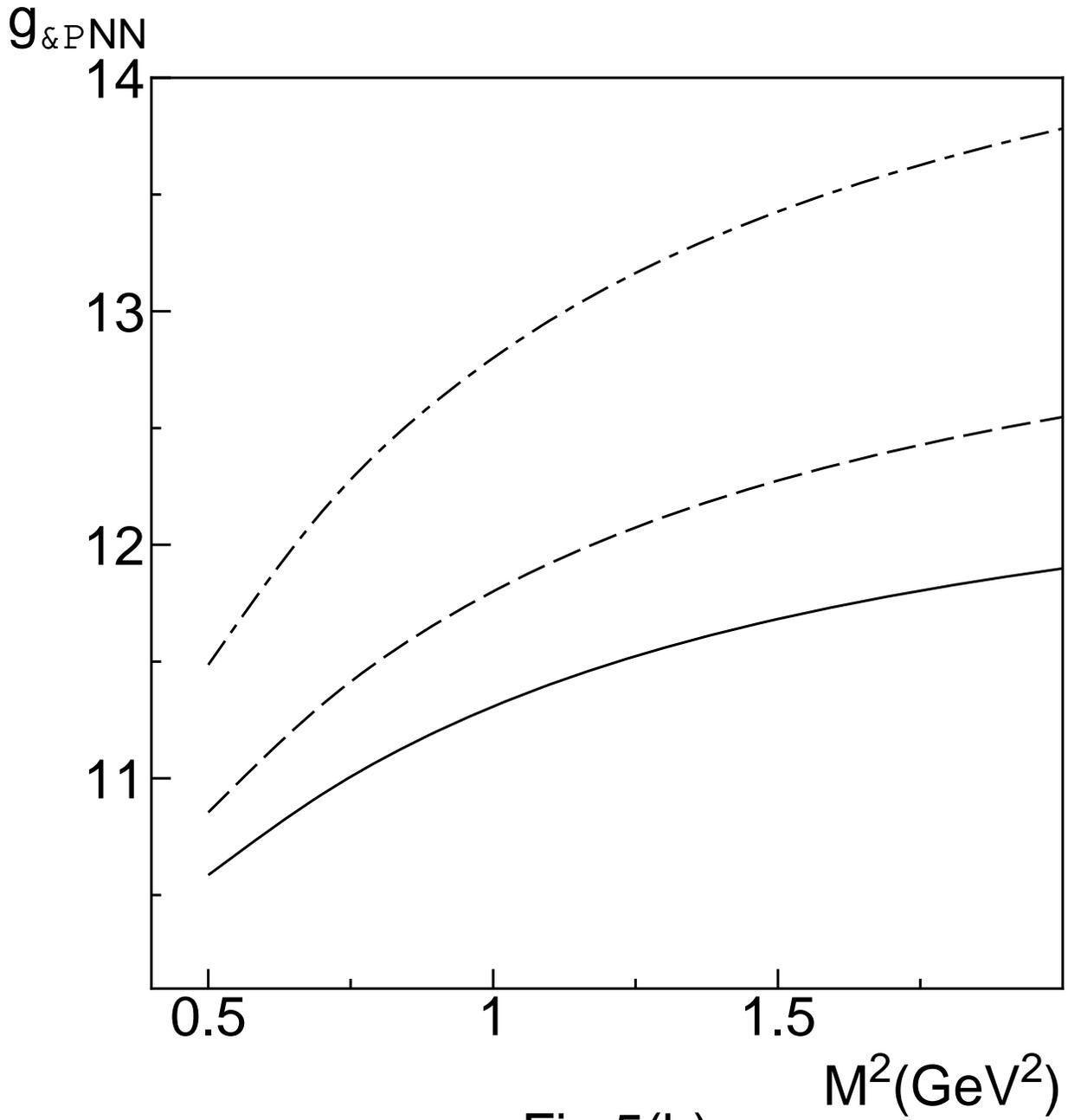
\end{document}